\documentclass[review,number,sort&compress]{elsarticle}

\usepackage{amsmath,amssymb}
\usepackage{textcomp}
\usepackage{color}
\usepackage{float}
\usepackage{subfig}
\usepackage[colorlinks, linkcolor={blue},citecolor={blue},breaklinks]{hyperref}

\journal{NIMA}

\begin{document}

\begin{frontmatter}

\title{The Double Spin Asymmetry of Nitrogen in Elastic and Quasielastic Kinematics from a Solid Ammonia Dynamically Polarized Target}

\author[HUJIaddress]{M. Friedman
\fnref{fn1}
\corref{mycorrespondingauthor}}
\fntext[fn1]{Current address: National Superconducting Cyclotron Laboratory, Michigan State University, East Lansing, Michigan 48824, USA}
\cortext[mycorrespondingauthor]{Corresponding author}
\ead{moshe.friedman@mail.huji.ac.il}
\author[DALHOUSIEaddress,JLABaddress,MARYSaddress]{J. Campbell}
\author[UVA]{D. Day}
\author[JLABaddress]{D.W. Higinbotham}
\author[MARYSaddress]{A. Sarty}
\author[HUJIaddress]{G. Ron}

\address[HUJIaddress]{Racah Institute of Physics, Hebrew University, Jerusalem, Israel 91904}
\address[DALHOUSIEaddress]{Dalhousie University, Halifax, Nova Scotia B3H 4R2, Canada}
\address[JLABaddress]{Thomas Jefferson National Accelerator Facility, Newport News, VA 23606, USA}
\address[MARYSaddress]{Saint Mary’s University, Halifax, Nova Scotia B3H 3C3, Canada}
\address[UVA]{University of Virginia, Charlottesville, VA, 22904, USA}

\begin{abstract}
Solid ammonia (NH$_3$) is commonly used as a dynamically polarized proton target for electron and muon scattering cross-section asymmetry measurements. As spin 1$^{+}$ particles, the $^{14}$N nuclei in the target are also polarized and contribute a non-trivial asymmetry background that should be addressed.  We describe here a method to extract the nitrogen contribution to the asymmetry, and report the cross-section asymmetries of electron-nitrogen scattering at beam energies of $E=1.7$ GeV and $E=2.2$ GeV, and momentum transfer of $Q^{2}=0.023-0.080$ GeV$^{2}$. 
\end{abstract}


\end{frontmatter}


\section{Introduction}

Scattering experiments of polarized lepton beams off polarized proton and other light targets has been used in the last few decades as a powerful tool for precision measurement of the nuclear electromagnetic form factors and spin structure functions (see, for example, Refs.~\cite{adeva1993,anthony1993,ackerstaff1997,anthony2000,goertz2002,warren2004,alexakhin2007,qian2011}). Solid ammonia, $^{14}$NH$_3$, is commonly used as a polarized proton target by exploiting the Dynamic Nuclear Polarization (DNP) technique to achieve proton polarizations over 90\%~\cite{meyer2004,crabb1995,crabb1997,averett1999}.

Previous works has shown that the spin 1$^{+}$ $^{14}$N is also polarized in the process to approximately 10\%, and hence contributes a nontrivial background to scattering asymmetry experiments \cite{court1986,adeva1998,rondon1999,kislev2013}. Specifically, experiments that utilize the asymmetry of the scattering cross section as a probe of the nucleon structure must take into account the asymmetry contribution by the polarized nitrogen nuclei.  However, direct experimental data of the background contribution to the measured asymmetry is not yet available. 

JLab experiment E08-007, GEp, measured the proton elastic form factor ratio, $\mu G_{E}/G_{M}$, at low momentum transfer, using elastic scattering of a polarized electron beam from a polarized $^{14}$NH$_3$ target~\cite{friedman2016}. In this report we present an analysis approach used to disentangle the nitrogen contribution to the asymmetry from the proton asymmetry. In addition, we report for the first time experimental cross section asymmetries of $^{14}$N elastic and quasi-elastic electron scattering.       

\section{Experimental Setup}
\begin{figure*}
  \centering
  \includegraphics[width=\linewidth]{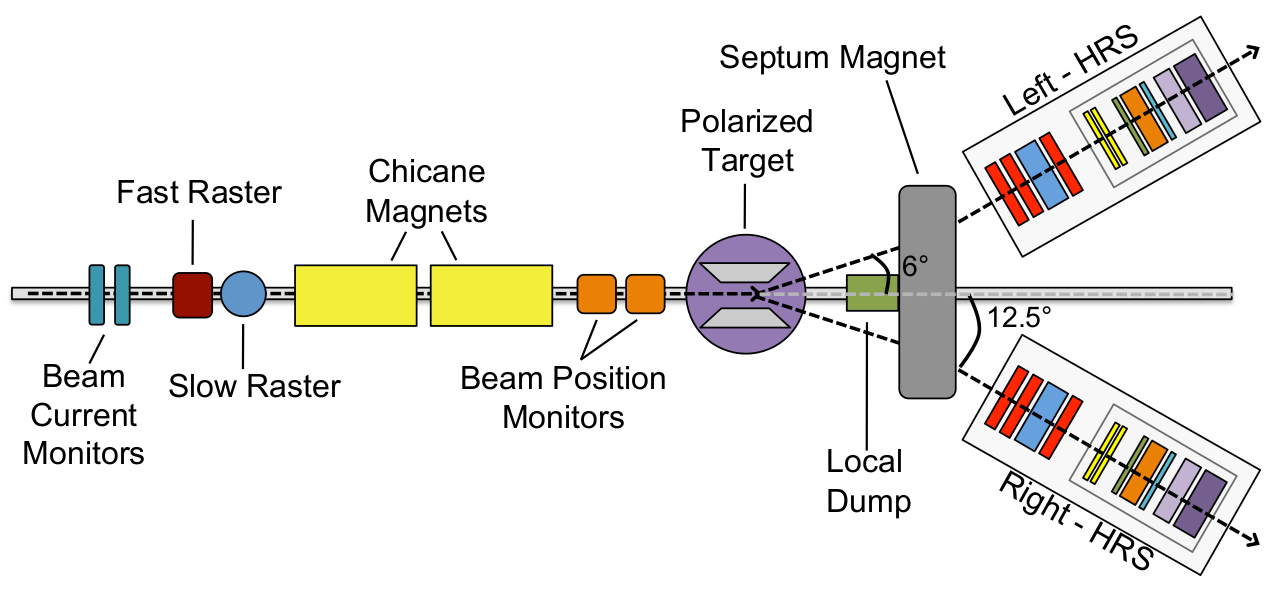}
  \caption{\label{fig:experimental_setup}Schematic diagram of the GEp experimental setup. See text for description. Figure from Ref. \cite{cummings2016}.}
\end{figure*}

The goal of the GEp experiment was to measure the proton elastic form factor ratio at a $Q^2$ range of 0.01-0.08 GeV$^{2}$, using the double spin asymmetry (DSA) technique~\cite{donnely1986}. The experiment was performed in Hall A of the Continuous Electron Beam Accelerator Facility (CEBAF) at the Thomas Jefferson National Accelerator Facility \cite{CEBAF,hallA}. A schematic view of the experimental setup is shown in Fig.~\ref{fig:experimental_setup}.
The CEBAF polarized beam at energies of 1.7 and 2.2 GeV, passed through fast and slow rasters and use two chicane magnets to compensate for the effect of the target magnetic field. The electrons, scattered off the polarized NH$_3$ target (Sec. \ref{sec:target}) and bent by the septum magnet, were detected by one of two High Resolution Spectrometers, HRSs. The beam current of $\sim$10 nA used for this experiment was too low to allow beam position monitoring \cite{friedman2016}. 

The target was polarized by a 5 T magnetic field, 5.6$^{\circ}$ left of the beam axis. The transverse component of the field causes downwards deflection of the beam. To compensate for this effect, two chicane magnets were installed 5.92 m and 2.66 m upstream of the target The two dipole magnets were tuned to direct the deflected beam towards the center of the target.

The data was taken at forward angles of $\sim 4^{\circ} - 7^{\circ}$. Due to technical limitations, the HRSs could not be placed at angles less than 12.5$^{\circ}$. For this reason, the target was placed 88 cm upstream of the traditional Hall A center, and two septum magnets were installed in front of the spectrometer to direct the scattered electrons from the forward angles to the actual HRSs position.

The data  was taken using the standard Hall A HRS spectrometers \cite{hallA}, and the analysis reported here is based on the left HRS data.
The beam polarization level was measured before, between, and after production runs of GEp, using the Hall A M{\o}ller polarimeter \cite{hallA,moller} (see Table \ref{tab:moller}). 
\begin{table}
   \centering
  \begin{tabular}{lc}
    \hline
    date & beam polarization \\
    \hline
    03/03/2012 &  -79.91 $\pm$ 0.20 \\
    03/30/2012 &  -80.43 $\pm$ 0.46 \\
    03/30/2012 &  +79.89 $\pm$ 0.58 \\  
    04/10/2012 &  -88.52 $\pm$ 0.30 \\  
    04/23/2012 &  +89.72 $\pm$ 0.29 \\  
    05/04/2012 &  -83.47 $\pm$ 0.57 \\  
    05/04/2012 &  -81.82 $\pm$ 0.59 \\  
    05/04/2012 &  +80.40 $\pm$ 0.45 \\  
    05/15/2012 &  +83.59 $\pm$ 0.31 \\  
    \hline
  \end{tabular}
  \caption{Beam polarization as measured by M{\o}ller polarimetry. Quoted uncertainties are statistical. Systematic uncertainties are estimated to be 1.7\% for all measurements. For details see \citep{moller_results}.}
  \label{tab:moller}
\end{table}

\subsection{\label{sec:target}Polarized NH$_3$ target}

A highly polarized proton target was needed for the GEp experiment. For this, we used the UVa solid NH$_3$ target~\citep{crabb1995,crabb1997,averett1999} that was successfully used for several experiments at JLab~\citep{warren2004,sane2009}.  The target operates at a temperature of $\sim$1 K, with a 5~T magnetic field. The protons are polarized via the Dynamic Nuclear Polarization (DNP) technique~\cite{overhauser1953,atsarkin2011}. For a detailed description of the target see \cite{keller2013,pierce2014} and references therein.

\subsubsection{Dynamic Nuclear Polarization}

Target polarization is defined as the difference between positive and negative aligned nuclear 
spins relative to the polarization axis, divided by the total number of nuclear spins:
\begin{equation}
  P=\frac{N_{\uparrow}-N_{\downarrow}}{N_{\uparrow}+N_{\downarrow}}.
\end{equation}
A traditional polarization technique is Thermal equilibrium (TE) polarization, also called “brute force polarization, where the target is cooled to low temperatures in a high magnetic field. In that case, the population of two magnetic sub levels is determined by the Boltzmann distribution:
\begin{equation}
  N_1 = N_2 \cdot \exp \left(-\frac{\Delta E}{k_B T} \right),
\end{equation}
where $\Delta E$ is the energy difference between the levels, $T$ is target temperature, $k_B$ is the Boltzmann constant, and $N_{1,2}$ are the number of nuclear spins in each sub level. For a spin-1/2 target, the polarization level obtained by TE is:
\begin{equation}
  P=\tanh \left( \frac{g \mu B}{2k_B T} \right). \label{eq:TE}
\end{equation}
Here $g$ is the particle g-factor, and $\mu$ is the nuclear magneton for protons or the Bohr magneton for electrons. Although for electrons TE polarization can reach above 90\% polarization, the significantly lower value of the proton magnetic moment makes TE much less effective for protons. For example, at a temperature of 0.5 K and a magnetic field of 5 T, the TE polarization of protons is 
about 1\%~\cite{atsarkin2011}.

Therefore the DNP technque is used to significantly increase nuclear polarization 
levels by applying microwave radiation to the target. In the UVa target, 
the increase in polarization is described by the equal spin 
temperature theory (EST). The spin temperature model is needed since 
the electron density is high in a solid NH$_3$ target 
and spin-spin interaction (SSI) plays an important role in the description of 
the system. The EST theory is summarized by Crabb and Meyer \cite{crabb1997}. In short, within the 
scope of the EST theory, each energy level contains a quasi-continuous band of spin-spin states. 
Those bands can be characterized by a different temperature than the environment by changing 
the occupations of the different sub-levels using microwave radiation.  The microwave 
radiation generates thermal mixing between the SSI temperature and nuclear Zeeman temperature, 
to achieve a much lower nuclear spin temperature. In practice, the target used for this 
experiment achieved polarization levels of between 70\% and 90\% under real experimental conditions. 

\subsubsection{Nitrogen Polarization}

In a work by B. Adeva \textit{et al.} \cite{adeva1998} the nitrogen asymmetry in a similar ammonia target was measured relative to the proton asymmetry and compared to a calculation based on the EST model. A substantial $^{14}$N polarization level of about 10\% was reported for the highly polarized target,in agreement with the EST model.

\section{Data Analysis}

\subsection{Target Polarization}

An NMR system was used for continuous measurement of the proton polarization level in the target. 
The NMR system used in this experiment is the same as the one used in previous experiments with 
this target. The signal from the NMR coil was connected to a Q-meter circuit to measure the
target polarization~\cite{court1993}.  A detailed analysis of the target proton polarization 
is reported by D. Keller~\cite{keller2013}. The nitrogen polarization level was calculated 
with respect to the proton polarization based on the EST model as calculated by Adeva \text{et al.}~\cite{adeva1998}.

\subsection{Optics}

The standard HRS optics reconstruction procedure described in Ref.~\cite{optics_general} 
is able to reconstruct the trajectories in cases where no target field is applied. 
In our case, a 5 T magnetic field around the target adds more complexity to the procedure. 
Therefore, the optics calibration and analysis is broken into two parts. 
The trajectories between the target and septa entrance are calculated 
by simulations of the electron motion in the magnetic field. The magnetic 
field is characterized by applying the Biot-Savart law to the current density distribution, and a cross-check is done by direct measurement of the target field. The uncertainty of the field map is less than 1.2\% over the whole region~\cite{chao2016}. The reconstruction of the trajectories from the entrance of the septa to the focal plane is done using an optics matrix. The calculation of the optics matrix is done using a sieve slit, and uses the well known behavior of elastic scattering and survey data described in\cite{survey1,survey2}. A simulation of the magnetic field is required for the optics matrix optimization, 
since linear propagation of the trajectories from the target to the sieve slit cannot be assumed. 

As part of the optics analysis a GEANT4 \cite{agostinelli2003} Monte-Carlo simulation of the experimental setup, g2psim, was developed~\cite{chao2016}. The simulation contains the materials and the field map along the trajectory of the scattered electron, and was used for the calibration of the optics matrix. The trajectories are calculated by integration of the equation of motion in the magnetic fields using the Runge-Kutta-Nystr\"{o}m method. Energy losses due to ionization, electron scattering, internal and external Bremsstrahlung are included in the simulation, as described in \cite{g2psim_eloss}.

The presence of the target field results in bending of the beam upstream of the target as well. 
To compensate for this effect, two chicane magnets were installed to lower then lift the beam trajectory
to achieve normal incidence of the electron beam on the center of the target.
However, even after this correction the beam showed deviations from the center and was slightly tilted relative to the beam axis. The tilt angle was calculated using BdL simulations. 

\begin{figure}
	\centering
	\includegraphics[width=\linewidth]{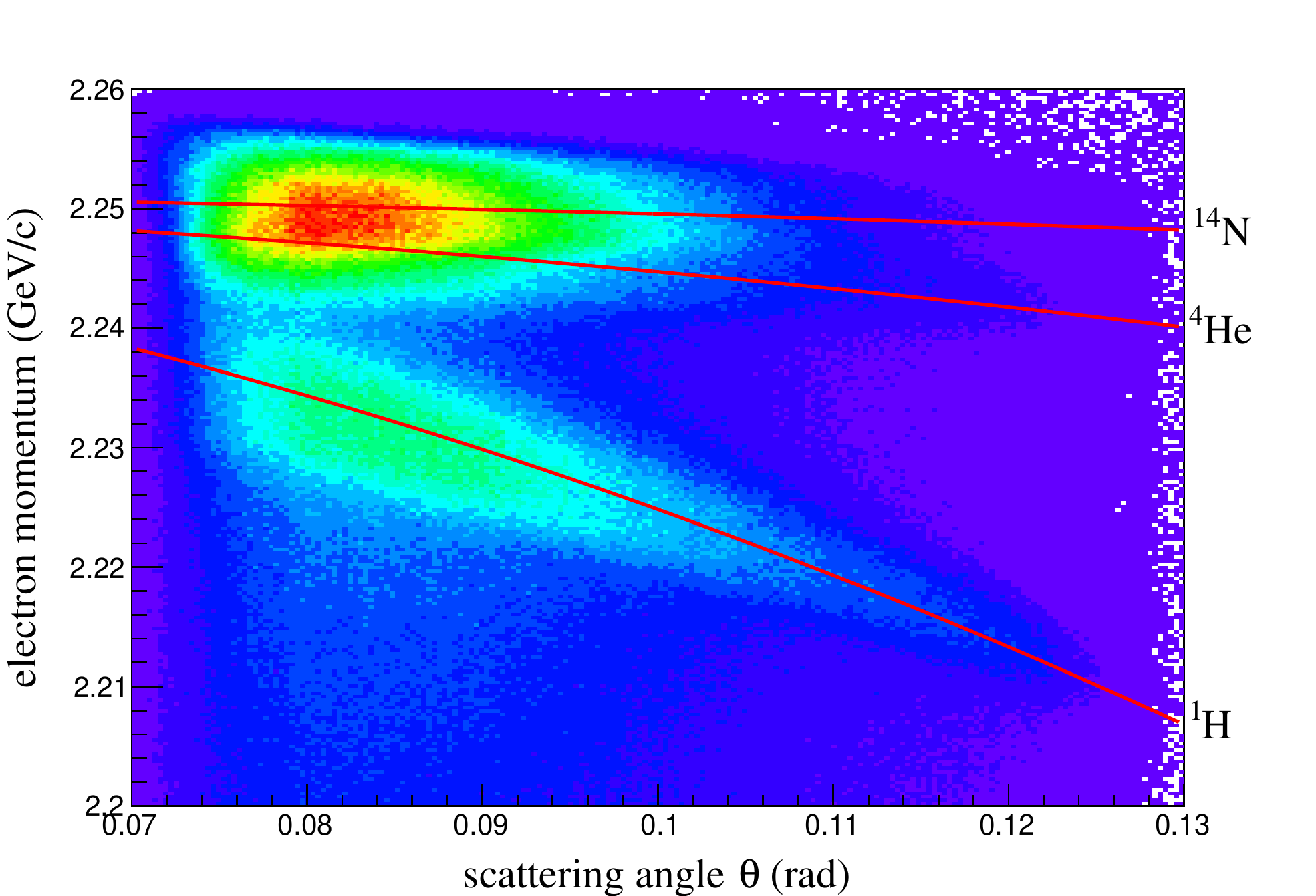}
	\caption{\label{fig:data_2D} (colour online) Typical distribution of reconstructed momentum and scattering angle variables. The separation between elastic scattering on hydrogen and the heavier elements $^{14}$N and $^4$He is evident, and the red curves compare the reconstructed variables to the elastic stripe formula (Eq. \ref{eq:elastic}).}
\end{figure}

Fig. \ref{fig:data_2D} shows an example of two dimensional histogram of scattered electron momentum vs. scattering angle. The elastic stripes for scattering off different nuclei (with mass M) are given by:
\begin{equation}
E^{\prime} = \frac{E}{1+\frac{E}{M}\left(1-\cos\theta \right)}, \label{eq:elastic}
\end{equation}
where $E^{\prime}$ is the scattered electron energy, $E$ is the incoming beam energy, and $\theta$ is the scattering angle. 
An imperfect scattering angle reconstruction results in small deviation of the variables in
Eq. \ref{eq:elastic} and is dealt with by applying an additional correction on the scattering angle. This correction is calculated as function of the reconstructed scattering angle by comparing to the expected scattering angle:
\begin{equation}
\cos\theta = 1 + M \left(\frac{1}{E}-\frac{1}{E^{\prime}} \right) \label{eq:elastic_reveresed} 
\end{equation}
around the proton elastic stripe.  In this way the elastic stripe can be used for fine tuning of the optics.

This uncertainty is estimated to be $\sim$1 MeV, based on the HRS resolution, electron energy loss approximation, and deviation of the invariant masses of hydrogen peaks from 0.9383 MeV. This translates into a $\sim$2 mrad systematic uncertainty in the scattering angle. 

\subsection{Asymmetry Extraction}

\begin{figure}
  \centering  
  \includegraphics[width=\linewidth]{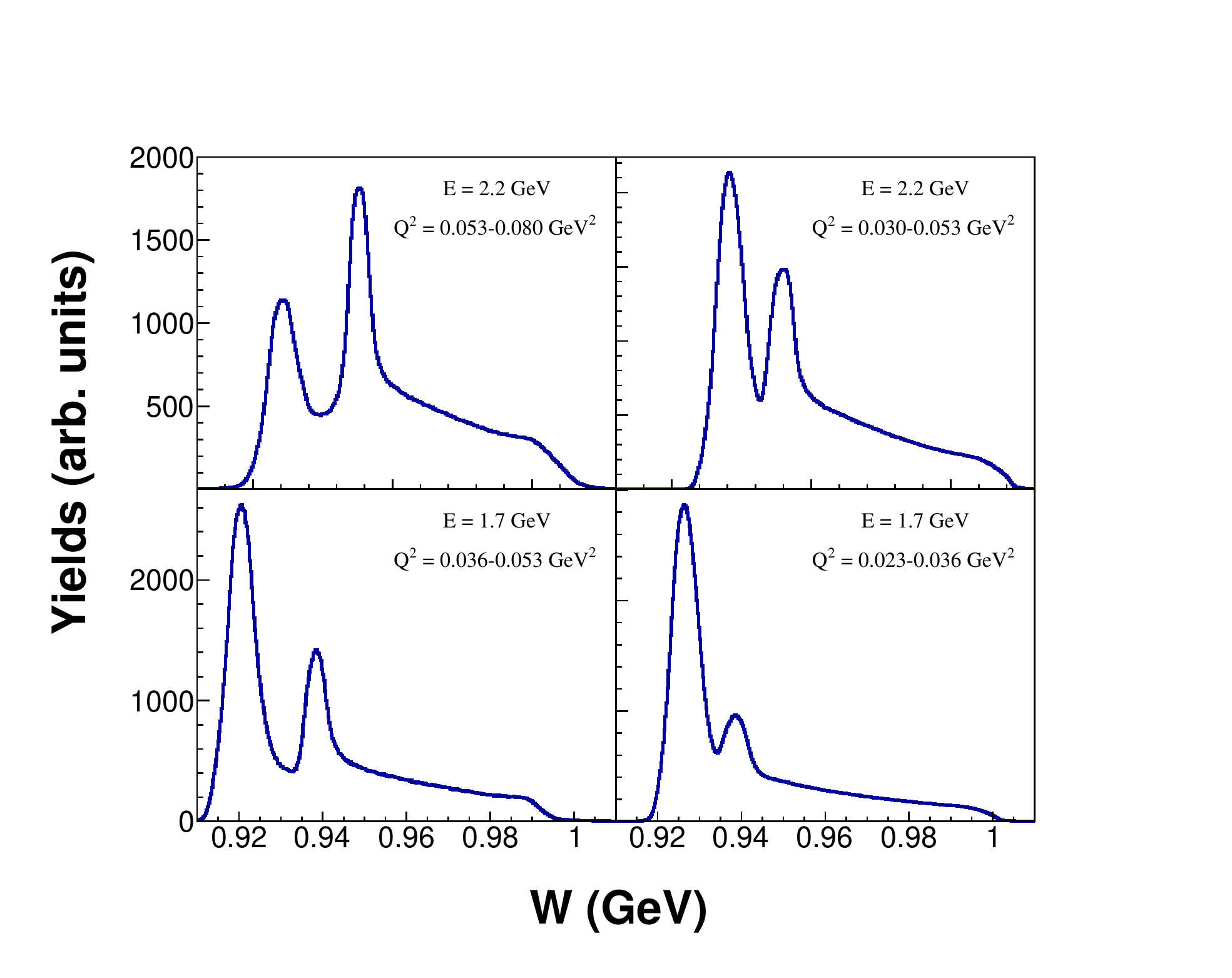}

\caption{\label{fig:W_plots}Invariant mass distributions for the left HRS at all experimental configurations. The hydrogen elastic events can be identified as the peak at $\sim$938 MeV.}
\end{figure}

Fig. \ref{fig:W_plots} shows the distribution of the invariant mass, $W$, for the different beam energy and momentum transfer values used for this analysis. The invariant mass is calculated using the target proton momentum $P^{\mu} = (M_p,0,0,0)$ and the momentum transfer $q^{\mu} = k^{\mu}-k^{\prime \mu}$, as:
\begin{equation}
  P_{tot}^{\mu} = q^{\mu}+P^{\mu},
\end{equation}
\begin{equation}
  W = \sqrt{P_{\mu,tot}P_{tot}^{\mu}}.
\end{equation}
The use of the invariant mass allows one-dimensional analysis that is equivalent to a two-dimensional analysis on momentum and scattering angle (Fig. \ref{fig:data_2D}). The invariant mass histograms are divided into three main regions: elastic scattering off $^{4}$He and $^{14}$N ($W\lesssim0.93$ GeV), elastic scattering off $^{1}$H ($W\sim0.938$ GeV), and quasi-elastic scattering off $^{4}$He and $^{14}$N ($W\gtrsim0.95$ GeV). Non-negligible mixing between the regions is due to the width of the elastic and quasi-elastic peaks, and due to radiative energy losses of the low $W$ scattering events. 

\begin{figure}
\centering
  \includegraphics[width=\linewidth]{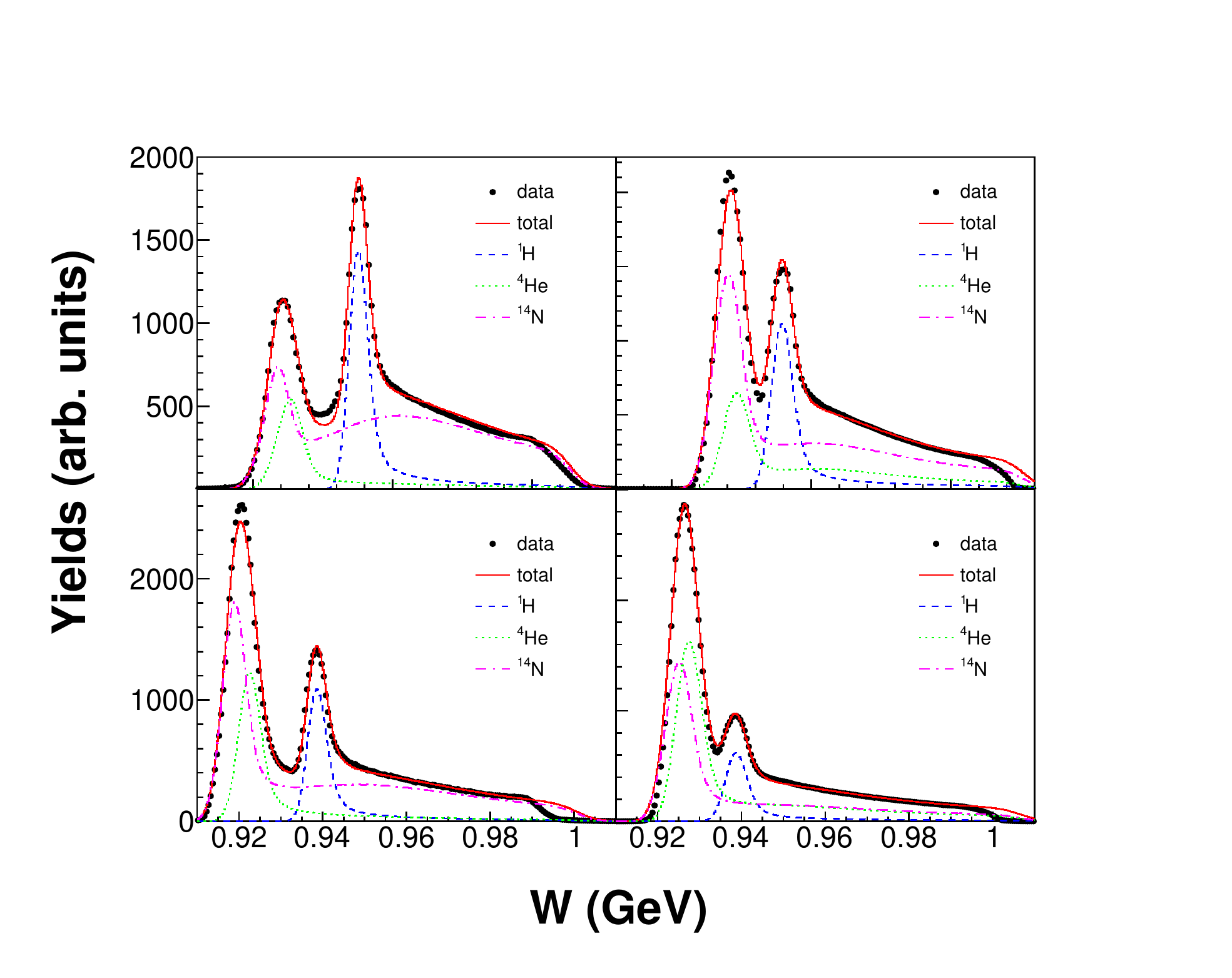}%

\caption{\label{fig:finalSim}Simulation results for all experimental configurations. The black circles are the experimental data, the red curves are the total simulated yields, the blue, green and purple curves are the proton, helium and nitrogen partial contributions, respectively.}
\end{figure}

In order to calculate the relative contribution of the different scattering components,we used g2psim to simulate the invariant mass distribution of each individual component. As an input to the simulation, elastic models for protons, $^4$He, and $^{14}$N were coded using experiment-based form factor parameterizations~\cite{arringtonFF,deJager}. As inelastic scattering data at the relevant kinematics are not available, g2psim uses two models, QFS~\cite{QFS} and PBosted~\cite{PBosted} as an input for the inelastic scattering simulation. The two models produced significantly different yields at the quasi-elastic region in our kinematics. Unpublished nitrogen data at similar kinematics were compared to these models, and the PBosted model was seen to better reproduce the quasi-elastic peak~\cite{zielinski2016}. This was also indicated by the quasi-elastic data collected in this experiment. Consequently, for this analysis, the PBosted model was selected. Gaussian smearing on the order of 1-2 MeV (1$\sigma$) and an energy shift on the order of 0.5-1 MeV was applied to the simulated data in order to account for inaccuracies in the simulated detection resolution and energy losses. The simulation was only used to produce the shape of the invariant mass distribution of the different scattering components. The relative amplitudes cannot be accurately derived from the cross section models, and the packing fraction (i.e., the mass ratio between helium and ammonia in the target) is only poorly known. Instead, the magnitude of each contribution is scaled to the data using a minimum $\chi^2$ fit.  Fig.~\ref{fig:finalSim} shows the simulation results for all experimental configurations after the correction and the fit. 

The above fitting procedure is not very sensitive to the differences between $^4$He and $^{14}$N elastic contributions, and it is almost completely insensitive to the differences between the quasi-elastic contributions of these two elements. Nevertheless, the differentiation between the three main regions in the spectrum, heavy element elastics, proton elastics, and quasi-elastics, is robust. Hence, we only use the combined $^{4}$He and $^{14}$N contributions for the analysis. We do not consider these fits reliable in terms of the differences between $^{4}$He and $^{14}$N yields. 

For an unpolarized background $B$, the experimental asymmetry (assuming 100\% beam and target polarization) is:
\begin{equation}
  A_{raw} = \frac{N^{+}-N^{-}}{N^{+}+N^{-}+B} = \frac{A_{phys}}{1+\frac{B}{N^{+}+N^{-}}},
\end{equation}
\begin{equation}
  A_{raw} = A_{phys}\cdot\left(1-\frac{B}{T}\right)
\end{equation}
where $A_{raw}$ is the raw asymmetry, $A_{phys}$ is the physical asymmetry, $N^{\pm}$ are the number of events with positive or negative helicities, and $T=N^{+}+N^{-}+B$ is the total event count. We define the dilution factor $f=1-B/T$ and add the corrections for the beam and target polarization to obtain:
\begin{equation}
  A_{phys} = \frac{A_{raw}}{fP_BP_T}, \label{eq:dilution}
\end{equation}
where $P_{B}$ and $P_{T}$ are beam and target polarization, respectively. In our case the polarized $^{14}$N contributes to the measured asymmetry, and Eq. \ref{eq:dilution} cannot be used. B. Adeva \textit{et al.} \cite{adeva1998} and O. A. Rondon \cite{rondon1999} used shell model approximations to determine the nitrogen asymmetry relative to the proton asymmetry. Here we propose a different approach, based on the measured asymmetries in the different regions of the invariant mass distribution. 
\begin{figure}
	\centering
	\includegraphics[width=\linewidth]{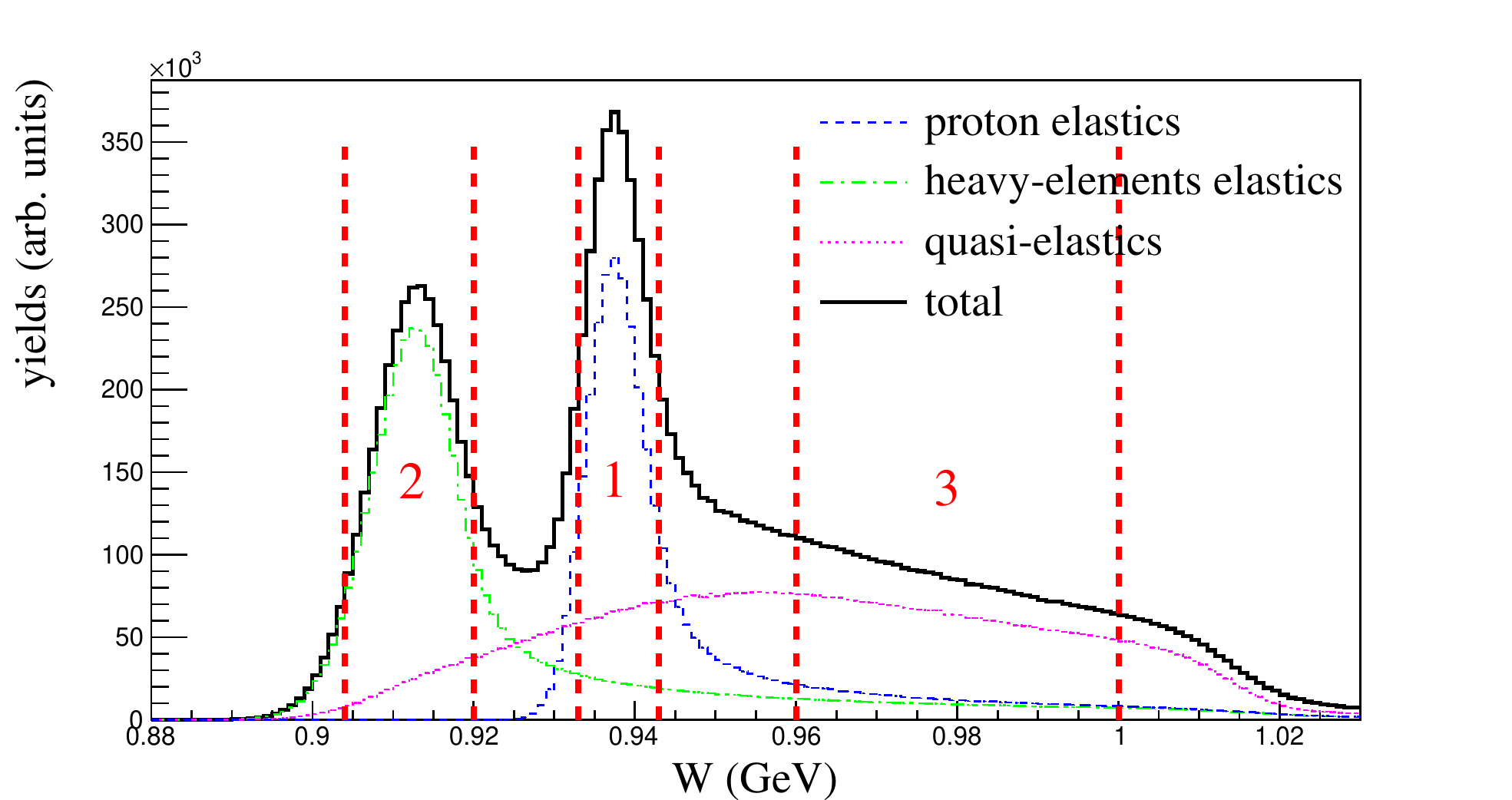}
	\caption{\label{fig:regions}An example of invariant mass histogram divided into three regions, each dominated by different reaction: proton elastics (1), heavy-elements elastics (2) and quasi-elastics (3). The partial contribution of each reaction is determined based on simulation.}
\end{figure}
The data is treated as if it consists of three reactions: proton elastics, heavy-element elastics, and quasi-elastics. To each of these reactions we attribute a physical asymmetry $A_1,A_2,A_3$, respectively. Each of these reactions dominates a different region in the invariant mass histogram, with some level of mixing of all three regions (see Fig. \ref{fig:regions}). Using the data we extract raw asymmetries for each region, $A_{raw,1},A_{raw,2},A_{raw,3}$. The raw asymmetries are related to the physical asymmetries by:
 \begin{equation}
 A_{raw,i} = \sum_{j=1}^3 C_{i,j}A_j, \label{eq:regions}
 \end{equation}
 where the coefficient $C_{i,j}$ is the partial contribution of reaction $j$ to the yields in region $i$, and is calculated from the simulation. This set of linear equations can be solved to extract the physical asymmetries for the three reactions $A_1,A_2,A_3$. The main advantage of this approach is that it is based on data, and does not require knowledge of the nitrogen polarization level and asymmetry, packing fraction, elastic or quasi-elastic absolute cross sections. It is noted, that the strengths of the three main regions are very clear in the data, and the successful reproduction of the quasi-elastic region by the simulation makes the determination of $C_{i,j}$ robust. A quantitative estimation of the uncertainties is detailed below. An additional important advantage is that we also extract, for the first time, physical asymmetries for $^{14}$N elastic and quasi-elastic scattering. Indeed, the unknown level of $^4$He background implies large relative uncertainties on the nitrogen asymmetries. 
 The statistical uncertainty of the raw asymmetries is given by:
 \begin{equation}
 \Delta A = \sqrt{\frac{4N^+ N^-}{\left(N^+ + N^- \right)^3}}. \label{eq:raw_asym_error}
 \end{equation}
 The statistical uncertainties of the physical asymmetries, $A_1,A_2,$ and $A_3$ were determined using
 a Monte-Carlo method. For each extraction of physical asymmetries, Eq. \ref{eq:regions} was solved $10^6$ times, with $A_{raw,i}$ randomly generated around the experimental values with the appropriate standard deviation. The mean of the resulting distribution was used as the physical asymmetry, and the standard deviation as the statistical uncertainty. The correlations between the three reactions were also calculated by Monte-Carlo using the definition:
 \begin{equation}
 \Sigma_{i,j} = \sum_n (X_{i,n}-\overline{X}_i)(X_{j,n}-\overline{X}_j)/N,
 \end{equation}
 where $X_{i,n}$ is the $n^{th}$ calculated asymmetry for reaction $i$, $\overline{X}_i$ is the mean asymmetry for reaction $i$, and $N$ is the total number of events in the Monte-Carlo analysis. Table \ref{tab:cov_matrices} shows the covariance matrices. Note the non-negligible off-diagonal elements. 
 
 \begin{table}
 	\centering
 	\subfloat[$0.053<Q^2<0.080$ GeV$^2$]{
 		\begin{tabular}{|l|ccc|}
 			\hline
 			$\times 10^{-7}$ & $A_1$ & $A_2$ & $A_3$ \\
 			\hline
 			$A_1$ & 4.66 & 0.21 & -1.57 \\
 			$A_2$ & 0.21 & 3.00 & -0.79 \\
 			$A_3$ & -1.57 & -0.79 & 3.07 \\
 			\hline  
 		\end{tabular}}
 		\quad
 \subfloat[$0.030<Q^2<0.053$ GeV$^2$]{
 \begin{tabular}{|l|ccc|}
 	\hline
 	$\times 10^{-7}$ & $A_1$ & $A_2$ & $A_3$ \\
 	\hline
 	$A_1$ & 1.92 & -0.06 & -0.64 \\
 	$A_2$ & -0.06 & 0.46 & -0.16 \\
 	$A_3$ & -0.64 & -0.16 & 1.36 \\
 	\hline  
 \end{tabular}}
 			
 \subfloat[$0.036<Q^2<0.053$ GeV$^2$]{
 \begin{tabular}{|l|ccc|}
   \hline
   $\times 10^{-7}$ & $A_1$ & $A_2$ & $A_3$ \\
   \hline
   $A_1$ & 6.62 & 0.01 & -1.93 \\
   $A_2$ & 0.01 & 1.03 & -0.48 \\
   $A_3$ & -1.93 & -0.48 & 4.49 \\
   \hline  
 \end{tabular}}
 \quad
 \subfloat[$0.023<Q^2<0.036$ GeV$^2$]{
 \begin{tabular}{|l|ccc|}
 	\hline
 	$\times 10^{-7}$ & $A_1$ & $A_2$ & $A_3$ \\
 	\hline
 	$A_1$ & 0.36 & -0.07 & -0.86 \\
 	$A_2$ & -0.07 & 0.24 & -0.16 \\
 	$A_3$ & -0.86 & -0.16 & 2.13 \\
 	\hline  
 \end{tabular}}
 					
 \caption{\label{tab:cov_matrices}Covariance matrices for 2.2 GeV (top) and 1.7 GeV (bottom) configurations. See text for the meaning of $A_1,A_2,A_3$. All values should be multiplied by $10^{-7}$.}  
\end{table} 				
The systematic uncertainties of the extraction were studied by changing the invariant mass cut 
width, and by changing the simulation energy and resolution calibrations in ranges that produce 
reasonable agreement with the data. The relative uncertainty on the nitrogen polarization 
is estimated to be 15\%, based on the differences between the EST model and the experimental 
data in~\cite{adeva1998}. Nitrogen elastic and quasi-elastic asymmetries are diluted 
by the presence of $^4$He. The ratio between helium and nitrogen in the data could 
not be extracted from the simulation. Based on packing fraction evaluation from JLab experiment E08-027 that 
ran parallel to this experiment using the same target \cite{cummings2016}, and general considerations, 
a conservative dilution factor of $f_{^{14}N}=0.75\pm0.25$ is used in Eq. \ref{eq:dilution} 
to extract the physical nitrogen asymmetries. 

\section{Results}

\begin{table}
\centering
  \begin{tabular}{ccccc}
    \hline
    E (GeV) & $Q^2$ (GeV$^2$) & $A$ (\%) & $\Delta A_{stat}$ (\%) & $\Delta A_{sys}$ (\%) \\
    \hline
    2.2 & 0.053-0.080 & 0.66 & 0.54 & 0.32 \\
    2.2 & 0.030-0.053 & 0.62 & 0.21 & 0.30 \\
    1.7 & 0.036-0.053 & 0.76 & 0.29 & 0.36 \\
    1.7 & 0.023-0.036 & 0.49 & 0.14 & 0.24 \\
    \hline
  \end{tabular}
  \caption{\label{tab:results_Nelastics}$^{14}$N elastic asymmetries with their absolute statistical and systematic uncertainties.} 
\end{table}

\begin{table}
\centering
  \begin{tabular}{ccccc}
    \hline
    E (GeV) & $Q^2$ (GeV$^2$) & $A$ (\%) & $\Delta A_{stat}$ (\%) & $\Delta A_{sys}$ (\%) \\
    \hline
    2.2 & 0.053-0.080 & -0.58 & 0.56 & 0.31 \\
    2.2 & 0.030-0.053 & 0.75 & 0.37 & 0.40 \\
    1.7 & 0.036-0.053 & -1.82 & 0.61 & 0.97 \\
    1.7 & 0.023-0.036 & -0.77 & 0.41 & 0.41 \\
    \hline
  \end{tabular}
  \caption{\label{tab:results_Nquasielastics}$^{14}$N quasi-elastic asymmetries with their absolute statistical and systematic uncertainties.} 
\end{table}

\begin{figure}[t]
\centering

\subfloat[$E=2.2$ GeV]{%
  \includegraphics[width=0.6\linewidth]{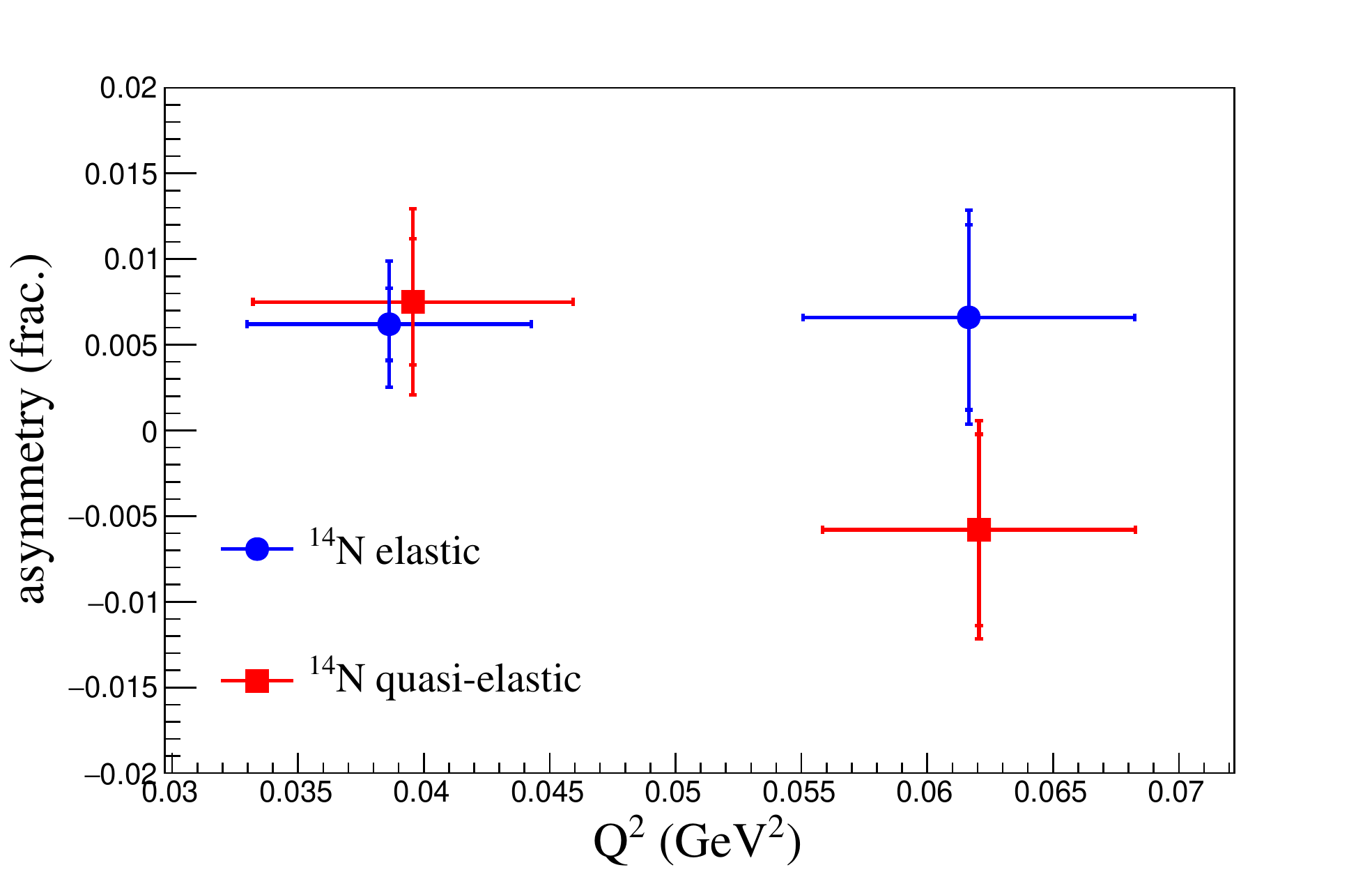}%
  \label{fig:Nresult22}%
}\qquad
\subfloat[$E=1.7$ GeV]{%
  \includegraphics[width=0.6\linewidth]{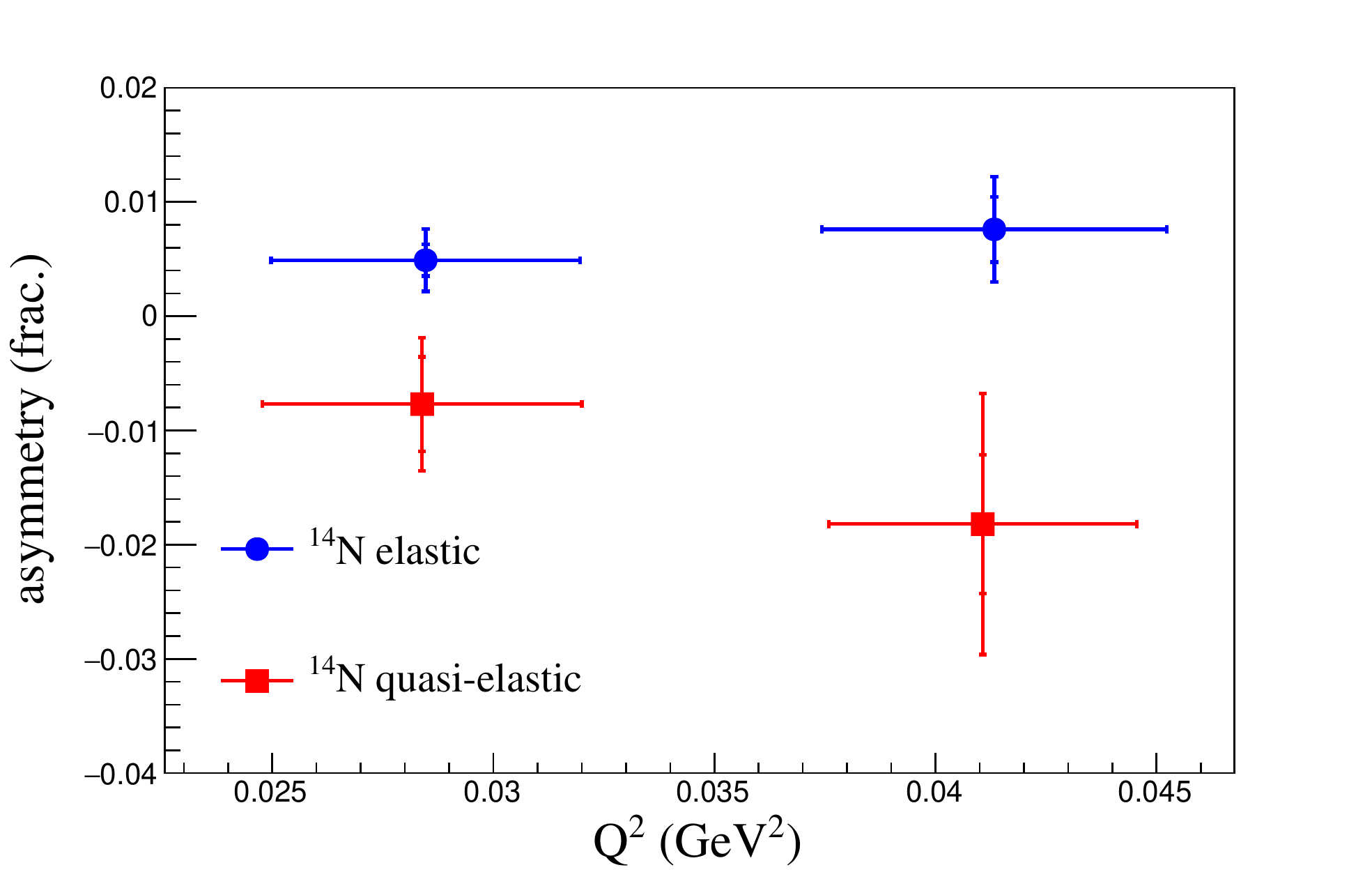}%
  \label{fig:Nresult17}%
}

\caption{\label{fig:Nresults}Physical asymmetries for electron-$^{14}$N elastic (blue circles) and quasi-elastic (red squares) scattering. The horizontal error bars represents the $Q^2$ range of which the data was collected.}
\end{figure}

\begin{table}
\centering
  \begin{tabular}{ccccc}
    \hline
    E (GeV) & $Q^2$ (GeV$^2$) & $A$ (\%) & $\Delta A_{stat}$ (\%) & $\Delta A_{sys}$ (\%) \\
    \hline
    2.2 & 0.053-0.080 & -3.57 & 0.061 & 0.142 \\
    2.2 & 0.030-0.053 & -2.41 & 0.044 & 0.096 \\
    1.7 & 0.036-0.053 & -2.73 & 0.078 & 0.109 \\
    1.7 & 0.023-0.036 & -2.02 & 0.019 & 0.081 \\
    \hline
  \end{tabular}
  \caption{\label{tab:results_proton}Proton asymmetries with their absolute statistical and systematic uncertainties.} 
\end{table}

Extracted asymmetries for electron-$^{14}$N elastic scattering are listed in Table \ref{tab:results_Nelastics} and asymmetries for electron-$^{14}$N quasi-elastics scattering are listed in Table \ref{tab:results_Nquasielastics}. Fig. \ref{fig:Nresults} compares both asymmetries. The extracted asymmetries for electron-proton elastics scattering are listed in Table \ref{tab:results_proton}. A summary of the uncertainties is given in Table \ref{tab:uncertainties}. The magnitude of the extracted asymmetries shows general agreement with the approximations made by B. Adeva \textit{et al.} \cite{adeva1998} and by O. A. Rondon \cite{rondon1999} in magnitude. However, the differences in sign between the elastic and quasi-elastic scattering requires further investigation. 

\begin{table}
\centering
  \begin{tabular}{ccccc}
    \hline 
    origin & statistical (\%) & systematic (\%) \\
    \hline
    beam polarization & 0.20 & 1.70 \\
    target polarization & 0.75 & 2.9 \\
    asymmetry cuts & - & 0-1.1 \\
    asymmetry extraction:\\
    nitrogen elastics & 30-80 & 10\\
    nitrogen quasi-elastics & 30-100 & 25\\
    proton elastics & 0.9-5.1 & 2.0-3.5\\
    $^4$He dilution & - & 44 \\
    \hline
  \end{tabular}
  \caption{\label{tab:uncertainties}A summary of asymmetry uncertainties for electron-$^{14}$N elastic, electron-$^{14}$N quasi-elastic, and electron-proton elastic scattering. $^4$He dilution is only relevant for the nitrogen-related asymmetries. The tables list ranges over the whole data-set, and specific total uncertainties are listed in Tables \ref{tab:results_Nelastics},\ref{tab:results_Nquasielastics},\ref{tab:results_proton}. } 
\end{table}

\section{Summary}

The nitrogen in dynamically-polarized NH$_{3}$ targets contributes a nontrivial asymmetry background for high-precision measurements of lepton-proton scattering cross-section asymmetry. We experimentally extracted the nitrogen contribution for the asymmetry at $Q^2$ value of 0.023-0.08 GeV$^{2}$. We also demonstrated a method for the extraction of these asymmetries that can be applied in similar experiments at different $Q^2$ values. The relative accuracy of our results is low, but sufficient to evaluate the absolute contribution of the nitrogen asymmetries for experiments with such targets.

\clearpage

\section*{Acknowledgments}

This work was supported by the U.S. Department  of  Energy contract  DE-AC05-06OR23177 under  which  Jefferson  Science  Associates operates  the Thomas Jefferson National Accelerator Facility.

\section*{References}
\bibliographystyle{elsarticle-num}
\bibliography{Friedman_Nasymmetry}

\end{document}